      [1999/12/01 v1.4c Il Nuovo Cimento]
\documentclass{cimento}

\newcommand       \msun        	{$M_{\odot}$}
\newcommand       \lsun      	{$L_{\odot}$} 
 
\newcommand      \misme     {\left<m_{\rm ISM}\right>}
\newcommand      \mism     {$\left<m_{\rm ISM}\right>$}
\newcommand      \spitzer     {{\it Spitzer}}
\newcommand 		\iso			{{\it ISO}}
\newcommand		\jay			{J1148+5251}
\newcommand      \yd      {$\widehat Y_d$}
\newcommand      \yde      {\widehat Y_d}
\newcommand 	     \mstar      {$m_{\star}$}
\newcommand 	     \mstare      {m_{\star}}

\newcommand      \rsne      {R_{\rm SN}}

\newcommand		\mug			{$\mu_g$}
\newcommand		\muge		{\mu_g}

\usepackage{graphicx}  
\title{The origin and evolution of dust in high-redshift  galaxies}
\author{Eli Dwek\from{ins:x}\ETC,
Fr\'ed\'eric Galliano\from{ins:y},
        \atque
Anthony P. Jones\from{ins:z}}
\instlist{\inst{ins:x} Observational Cosmology Laboratory, Code 665, NASA Goddard Space Flight Center, \\Greenbelt, MD 20771, U.S.A.,
\inst{ins:y} Department of Astronomy, University of Maryland,  College Park, MD 20742, U.S.A.,
  \inst{ins:z} Institut d'Astrophysique Spatiale, F-91405 Orsay,  France} 
\PACSes{\PACSit{95.30.Wi, 98.58.Ca, 98.58.Jq, 98.62.Bj}
\PACSit{---.---}{\ldots}}
\begin{document}

\maketitle

\begin{abstract}
Dusty hyperluminous galaxies in the early universe provide unique environments for studying the role of massive stars in the formation and destruction of dust.
At redshifts above ~ 6, when the universe was less than ~ 1 Gyr old, dust could have only condensed in the explosive ejecta of Type~II supernovae (SNe), since most of the progenitors of the AGB stars, the major alternative source of interstellar dust, did not have time to evolve off the main sequence. We present analytical models for the evolution of the gas, dust, and metals in high redshift galaxies, with a special application to SDSS J1148+5251, a hyperluminous quasar at $z = 6.4$ hereafter referred to as \jay. We show that an average SN must condense at least 1~\msun\ of dust to account for the mass of dust in this object, when grain destruction by supernova remnants (SNRs) is taken into account. This required yield is in excess of $\sim 0.05$~\msun, the largest mass of dust inferred from infrared observations of Cas~A. If the yield of Cas~A is typical, then other processes, such as accretion onto preexisting grains in molecular clouds are needed to produce the mass of dust in J1148+5251. For such process to be effective, SNR must significantly increase, presumably by non-evaporative grain-grain collisions during the late stages of their evolution, the number of nucleation centers onto which refractory elements can condense in molecular clouds. 
\end{abstract}

\section{The presence of massive amounts of dust at high redshift}

\begin{figure}[htbp]
  \begin{center}
\includegraphics[width=4.0in]{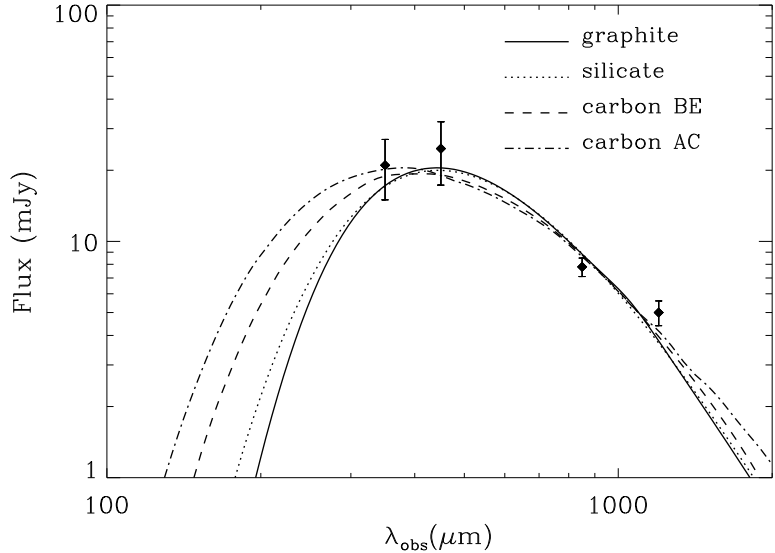}  
\end{center}
 \caption{Spectral fits of several possible dust compositions to the observed far-IR and submillimeter observations of \jay. Wavelengths are in the observer's frame of reference. Dust masses range from $(1-5)\times10^8$~\msun\ and luminosities range from $(2-3)\times 10^{13}$~\lsun, depending on dust composition. References to the observations can be found in \cite{ref:dwek07}.}
    \label{dustspec}
\end{figure}

The detection of massive amounts of dust in hyperluminous infrared (IR) galaxies at redshifts $z > 6$ raises challenging questions about the sources capable of producing such large amount of dust during the relatively short lifetime of these galaxies (e.g. \cite{ref:maiolino06,ref:beelen06}). 
For example, the galaxy SDSS J1148+5251 (hereafter \jay) located at $z = 6.4$ was observed at far-IR and submillimeter wavelength \cite{ref:bertoldi03,ref:robson04,ref:beelen06}. Figure \ref{dustspec} depicts the observed fluxes and spectral fits of various dust emission models through the data. The average IR luminosity of the source is $L_{IR} \sim 2\times 10^{13}$~\lsun, and the average dust mass is $M_d \sim 2\times 10^8$~\msun. Using the Kennicutt relation \cite{ref:kennicutt98}, one can derive a star formation rate (SFR) of $\sim 3000$~\msun~yr$^{-1}$ from the observed far-IR luminosity. For comparison, the Milky  Way galaxy is about 10~Gyr old, has an average SFR of $\sim 3$~\msun~yr$^{-1}$, and contains about $5\times 10^7$~\msun\ of dust, a significant  fraction of which  was produced in AGB stars. 

At $z=6.4$ the universe was only 890~Myr old, using standard $\Lambda$CDM parameters ($\Omega_m = 0.27$, $\Omega_{\Lambda} = 0.73$, and $H_0 = 70$~km~s$^{-1}$~Mpc$^{-1}$). If \jay\ formed at $z = 10$ then the galaxy is only 400~Myr old.
If the SFR had occured at  a constant rate over the lifetime of the galaxy, its initial mass should have been about 10$^{12}$~\msun, which is significantly larger than the dynamical mass $M_{dyn} \approx 5\times 10^{10}$~\msun\ of the galaxy \cite{ref:walter04}. The high observed SFR may therefore represent a recent burst of star formation that has lasted for only about 20~Myr. 
The galaxy \jay\ is therefore at most $\sim 400$~Myr old, and probably significantly younger with an age of only $\sim 20$~Myr. Adopting a current gas mass of $M_g = 3\times 10^{10}$~\msun\ for this galaxy we get that the gas mass fraction at 400~Myr is about 0.60. The dust-to-gas mass ratio is given by $Z_d \equiv M_d/M_g = 0.0067$. 

A significant  fraction of the dust in the Milky Way was produced in AGB stars. However, these stars are not likely  to  contribute significantly to the formation of dust in very young galaxies, since the low mass stars ($M \approx 2$~\msun) that  produce most of the dust did not have time to evolve off the main sequence \cite{ref:dwek98,ref:morgan03,ref:dwek07}.  
In contrast, core collapse supernovae (SNe) ($M > 8$~\msun) inject their nucleosynthetic products back into the ISM shortly ($t < 20$~Myr) after their formation, resulting in the rapid enrichment of the interstellar medium (ISM) with the dust that formed in their explosive ejecta. But can SNe account for the large amount of dust seen in this object? The answer to this question is complicated by the fact that SNe are also the main source of grain destruction during the remnant phase of their evolution \cite{ref:jones96,ref:jones04}. The problem can therefore only be addressed with a detailed model for the evolution of dust in these systems.

\section{The evolution of dust in evolved galaxies}
The evolution of dust is best studied in the framework of chemical evolution models that follow the galactic metal enrichment resulting from stellar nucleosynthesis (e.g. \cite{ref:matteucci97}). Such chemical evolution models need to be generalized to include processes unique to the evolution of dust: the condensation efficiency of the refractory elements in stellar ejecta, the destruction of grains in the ISM by expanding supernova remnants (SNRs), and the growth and coagulation of grains in clouds \cite{ref:dwek98}.

  \begin{figure}[htbp]
  \begin{center}
\includegraphics[width=5.0in,trim=0 -1.0in 0 0]{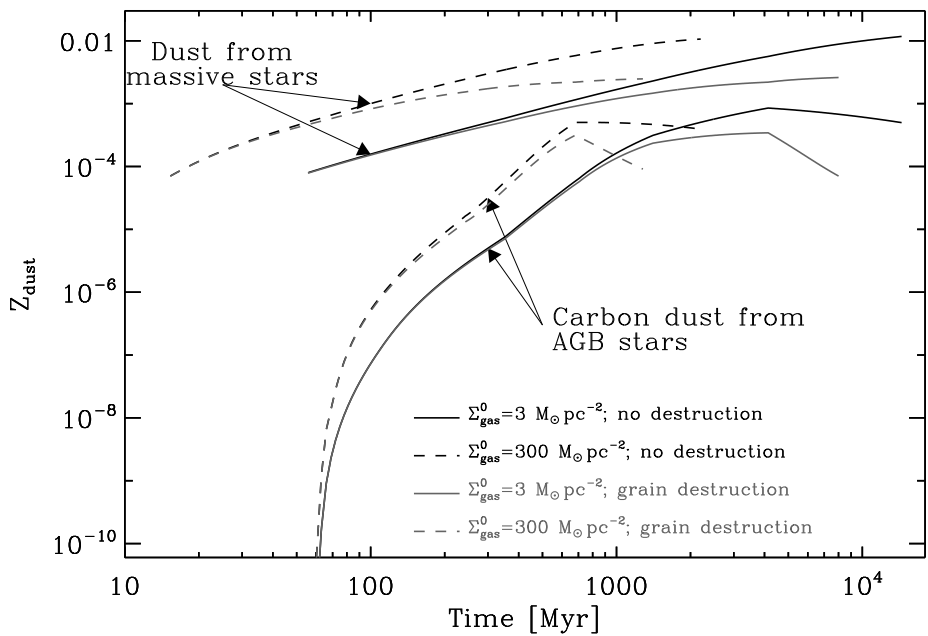}  
\end{center}
  \vspace{-0.7in}
 \caption{The evolution of SN-condensed silicate and carbon dust and AGB-condensed carbon dust as a function of time. The model used to produce the figure is described in the text, and is presented in more detail in \cite{ref:galliano07}.}
   \label{dustvol}
\end{figure}

Figure \ref{dustvol} shows the evolution of SN-- and AGB--condensed dust as a function of time. Details of the dust evolution model are given in \cite{ref:galliano07}. SNe can form both silicate and carbon dust since their ejecta have layers with different C/O ratios that may be larger or smaller than unity which are never microscopically-mixed. In contrast, the ejecta of AGB stars are either carbon- or oxygen-rich. Ideally, AGB stars are therefore expected to form either carbon or silicate dust. Observations show that some stellar outflows may have more complex chemistry which may be the results of mass loss from a binary system, since they show the presence of both silicates and C-rich dust products \cite{ref:waters04}. For simplicity, the evolutionary models used in Figure \ref{dustvol} assume that stars with masses below 8~\msun\ form either carbon or silicate dust depending on the C/O ratio in their ejecta, and that higher mass stars that end as core--collapse SNe form both silicate and carbon dust according to the prescription used by \cite{ref:dwek98}. For AGB stars, the figure depicts only their contribution to the carbon dust enrichment of the ISM. The most massive AGB stars are O-rich, so that the onset of the AGB contribution to the  carbon dust abundance is delayed until the C-rich stars evolve off the main sequence. However, the IMF-weighted yield of carbon dust peaks at stellar masses of about 2~\msun, so most of the carbon dust is injected into the ISM with a delay time between 0.4 and 4~Gyr, depending on the star formation history and the grain destruction rate. The subsequent decline in the yield of carbon dust results from the fact that lower mass AGB stars are O-rich and will only form silicates in their ejecta.      

  \begin{figure}[htbp]
  \begin{center}
\includegraphics[width=5.0in]{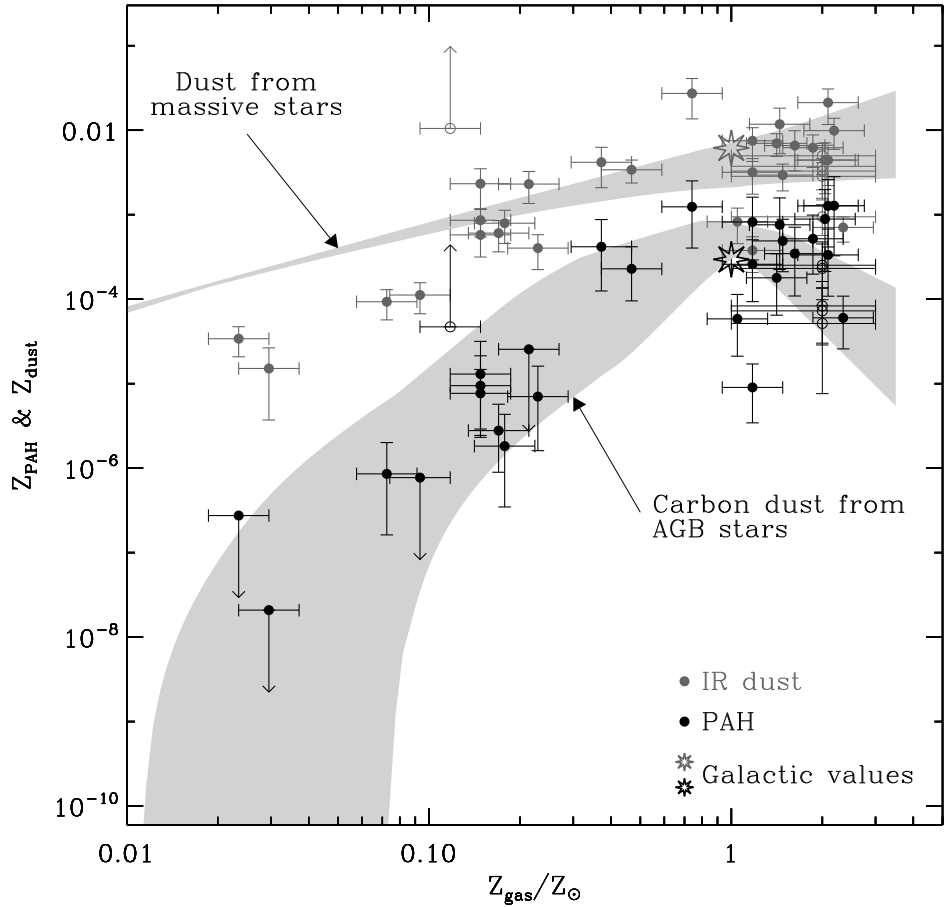}  
\end{center}
 \caption{ Comparison between the metallicity trends of the PAH abundance derived from the observed SED and those derived from the 
           chemical evolution model.
           The shaded area represent the range of model prediction for different grain destruction and star formation rates. Details of the figure are described in \cite{ref:galliano07}.}
   \label{pahvol}
\end{figure}

The qualitative features of this dust evolution model were recently confirmed by mid-IR photometric and spectral observations of the features of polycyclic aromatic hydrocarbon (PAH) molecules in nearby galaxies with the \spitzer\ and \iso\ {\it Infrared Space Observatory} satellites (see \cite{ref:galliano07} for references). The observations showed a deficiency in PAH emission from low metallicity galaxies, and a trend of increased emission in galaxies with higher metallicity. The PAH {\it abundance}, derived from detailed modeling of the spectral energy distribution of nearby galaxies \cite{ref:galliano07}, also correlated with their metallicity. This trend can be explained as a natural consequence of stellar evolutionary effects. Since PAHs are made in AGB stars, the trend simply reflects the delayed injection of these macromolecules into the ISM \cite{ref:dwek05,ref:galliano07}. Figure \ref{pahvol} is a mapping of the trend of PAH abundance with age onto a trend with metallicity for the same chemical evolution model. The figure illustrates the disctinct evolutionary trends of SN- and AGB-condensed dust. The shaded area represents the range of model predictions for different grain destruction and star formation rates. Details of the chemical evolution model can be found in \cite{ref:galliano07}. 

\section{The chemical evolution of dust in young systems}
The detailed dust evolution models described above show that the contribution of AGB stars to metal and dust abundance can  be neglected in galaxies with ages less than about 400~Myr. 
The equations for the chemical evolution of the galaxy can then be solved using the instantaneous recycling approximation, which assumes that stars return their ejecta back to the ISM promply after their formation. In the following we present the results for a closed box model. The case for the infall model is given in \cite{ref:dwek07}.

In the closed box model, the mass of the ISM changes due to the process of star formation (astration) at a rate given by:
\begin{equation}
\label{ }
{dM_g\over dt} = -(1-R) \psi(t)
\end{equation}
where $R$ is the fraction of the stellar mass that is returned back to the ISM during the stellar lifetime.
Assuming that  the SFR, $\psi(t)$, is proportional to the gas mass to some power $k$, we can write that:
\begin{eqnarray}
\psi(t) & =  & \psi_0\ \left({M_g(t)\over M_0}\right)^k \\ \nonumber
 & = &  \psi_0\ \muge(t)^k
\end{eqnarray}
where $M_0$ is the mass of the system which initially consists of only gas, \mug(t) is the fractional mass of gas at later times $t$, and $\psi_0$ is a parameter that is adjusted to fit the observed gas mass fraction at $t = 400$~Myr, taken to be the age of the galaxy.  

The equation for $dM_g/dt$ can then be written in terms of the gas mass fraction as:
\begin{equation}
{d\muge(t)\over dt} = -(1-R)\ \left({\psi_0\over M_0}\right)\ \muge(t)^k
\label{dmugdt}
\end{equation}

Analytical solutions are available for arbitrary values of $k$:
\begin{eqnarray}
\muge(t) & = &  \ \exp\left[ -(1-R)\ \left({\psi_0\over M_0}\right) t \right] \qquad \qquad \qquad k=1 \\ \nonumber
 & = & \left[ 1 -(1-R)(1-k)\ \left({\psi_0\over M_0}\right) t \right]^{1\over 1-k} \qquad \ \ k \neq 1
 \label{mug}
\end{eqnarray}
with the initial condition that \mug\ = 1 at time $t=0$.

The equation for the evolution of the mass of dust, $M_d$ in the  galaxy is given by:
\begin{equation}
{dM_d(t)\over dt} = -Z_d\ \psi(t) + \yde R_{SN}(t)- {M_d(t)\over \tau_d} \ .
 \label{dmddt}
\end{equation}
The first term represents the destruction rate of dust  by astration, the second term its  formation rate by SNe, and the third term the destruction rate of the dust by expanding SN blast waves. In the equation above, $Z_d \equiv M_d/M_g$ is the dust-to-gas mass ratio, \yd\ is the average yield of dust in Type~II supernovae, $R_{SN}$ is the galaxy's supernova rate determined by the SFR and the stellar IMF, and $\tau_d$ is the lifetime of the dust against destruction by SNRs. The lifetime for grain destruction is given by:
\begin{equation}
\tau_d = {M_g(t)\over \misme R_{SN}}
\label{taud}
\end{equation}
where \mism\ is the effective mass of ISM gas in which the dust grains are completely destroyed by a single SNR.
In our Galaxy, $M_g \approx 5\times10^9$~\msun, $R_{SN} \approx 0.02$~yr$^{-1}$ and \mism\ $\approx 500$~\msun, giving an average dust lifetime of $\approx 5\times 10^8$~yr \cite{ref:jones96}.
The SN rate can be written as:
\begin{equation}
\label{ }
\rsne = {\psi(t) \over \mstare}
\end{equation}
where \mstar\ is the mass of all stars born per supernova event. For example, for a Salpeter IMF in the 0.1 and 100~\msun\ range of stellar masses, \mstar $= 147$~\msun. 
 
Using equations (\ref{dmugdt}) and (\ref{dmddt}), the evolution of the dust mass can be expressed as a function of the remaining gas fraction as:
\begin{equation}
M_d(\muge)  =  \yde\ \left[{M_g(\muge)\over \misme + R\ \mstare}\right]\ \left(1-\muge^{\nu-1}\right) \label{md_mu} 
\label{md_mu}
\end{equation}
where $\nu$ is a dimensionless parameter:
\begin{equation}
\nu \equiv {\misme + \mstare \over (1-R)\ \mstare}
\label{nu_eq}
\end{equation}
The expression for $M_d(\muge)$ is independent of the star formation history of the  galaxy, but depends on the details of the IMF.

Equation (\ref{md_mu}) can be rewritten to give the dust yield required to obtain a given dust-to-gas mass ratio, $Z_d \equiv M_d/M_g$,  when the galaxy reaches a given gas mass fraction \mug:
\begin{equation}
\yde =Z_d\ \left[{\misme + R\ \mstare \over 1-\muge^{\nu-1}}\right]
\label{ydzd}
\end{equation}

\begin{figure}
    \begin{center}
\includegraphics[width=5.0in]{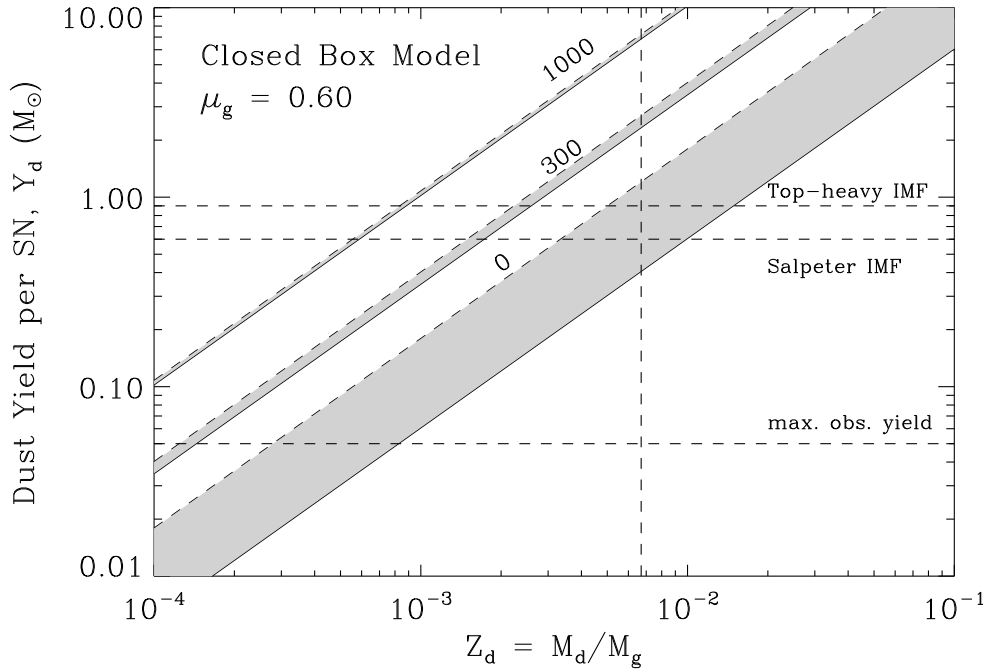} 
  \end{center} 
 \caption{The IMF-averaged yield of dust by type~II supernova, \yd, that is required to account for a given dust-to-gas mass ratio $Z_d$, is presented for different values of \mism\ given in units of \msun. Solid and dashed lines correspond to calculations done for a top-heavy and a Salpeter IMF, respectively. The horizontal dashed line near the bottom of the figure corresponds a value of $Y_d = 0.054$~\msun, the highest inferred yield of dust in supernova ejecta to date \cite{ref:rho07}. The vertical dotted line represents the value of $Z_d$ at $\muge =0.60$. Curves are labeled by \mism\ given  in units of \msun.  The top two dashed (solid) horizontal lines represent IMF-averaged theoretical dust yields for a Salpeter (top-heavy) IMF. }
   \label{snyield}
\end{figure}

Figure~\ref{snyield} shows how much dust an average SN {\it must} produce in order to give rise to a given dust-to-gas mass ratio, for various grain destruction efficiencies. The value of \yd\ was calculated when \mug\ reaches a value of 0.60, the adopted gas mass fraction of \jay\ at 400~Myr. 
Calculations were performed for two different functional forms of the stellar IMF: a Salpeter IMF in which $\phi(m) \sim m^{-2.35}$ and $0.1 < m$(\msun) $< 100$; and a top heavy IMF characterized by the same mass limits but a flatter slope $\phi(m) \sim m^{-1.50}$. Here, $\phi(m)$ is the number of stars per unit mass interval,  normalized to unity between 0.1 and 100~\msun.

The figure shows that, for example, to produce a value of $Z_d = 0.0067$ at \mug\ = 0.60, a SN must produce about 0.4~(1.2)~\msun\ of dust for a top-heavy (Salpeter) IMF, provided the dust is not destroyed in the ISM, that is, \mism\ = 0. Even with modest amount of grain destruction, \mism\ = 100~\msun, the required SN dust yield is dramatically increased to about $1-2$~\msun, depending on the IMF. 
The horizontal line in the figure corresponds to a value of \yd\ = 0.054~\msun, the largest mass of dust inferred to be present in the ejecta of the supernova remnant Cas~A \cite{ref:rho07}. Contrary to the claim by \cite{ref:rho07}, this yield is not sufficient to account for the large amount of dust observed in high redshift galaxies, since the chemical evolution models of \cite{ref:morgan03} do not include the effect of grain destruction. The figure shows that even without grain destruction, the largest observed yield can only give rise to a dust-to-gas mass ratio of $\sim 4\times10^{-4}$. If the mass of dust in the ejecta of Cas~A represents a typical SN yield, then other processes, such as accretion onto preexisting grains in molecular clouds is needed to produce the mass of dust in J1148+5251.

\section{Summary}
The early universe is a unique environment for studying the role of massive stars in the formation and destruction of dust. 
 The equations describing their chemical evolution can be greatly simplified by using the instantaneous recycling approximation, and by neglecting the delayed contribution of low mass stars to the metal and dust abundance of the ISM. Neglecting any accretion of metals onto pre-existing dust in the interstellar medium, the evolution of the dust is then primarily determined by the condensation efficiency of refractory elements in the ejecta of Type~II supernovae, and the destruction efficiency of dust by SN blast waves.

We applied our general results to \jay, a dusty, hyperluminous quasar at redshift $z = 6.4$ and found that the formation of a dust mass fraction of $Z_d = 0.0067$ in a galaxy with an ISM mass of $3\times 10^{10}$~\msun, requires an average SN to produce between 0.5 and 1~\msun\ of dust if there was no grain destruction (Fig. \ref{snyield}). Such large amount of dust can be produced if if the condensation efficiency in SNe is about unity. Observationally, the required dust yield is in excess of the largest amount of dust ($\sim 0.054$~\msun) observed so far to have formed in a SN. This suggests that accretion in the ISM may play an important role in the growth of dust mass.
For this process to be effective, SNR must significantly increase, presumably by non-evaporative grain-grain collisions during the late stages of their evolution, the number of nucleation centers onto which refractory elements can condense in molecular clouds.

\acknowledgments
This work was supported by NASA's LTSA 03-0000-065. The work of F.G. was supported by Research Associateship awards from the National Research Council (NRC) and from the Oak Ridge Associated Universities (ORAU) at NASA Goddard Space Flight Center.

\end{document}